\documentclass[prb,aps,twocolumn,superscriptaddress,showpacs]{revtex4}
\usepackage{amsfonts,amsmath,bm}

\usepackage{graphicx}

\newcommand{\bk}{b_{\bm{k}}}

\newcommand{\bkd}{b_{\bm{k}}^{\dag}}

\newcommand{\bmkd}{b_{-\bm{k}}^{\dag}}
\newcommand{\sumk}{\sum_{\bm{k}}}

\newcommand{\wk}{\omega_{\bm{k}}}

\newcommand{\gk}{g_{\bm{k}}}

\newcommand{\kk}{\bm{k}}

\newcommand{\rl}{\rangle\!\langle}

\DeclareMathOperator{\tr}{Tr}

\begin{document}

\title{Dephasing in the adiabatic rapid passage in quantum dots: the
  role of phonon-assisted biexciton generation}  

\author{Krzysztof Gawarecki}
\affiliation{Institute of Physics, Wroc{\l}aw University of
Technology, 50-370 Wroc{\l}aw, Poland}
\author{Sebastian L\"uker}
\affiliation{Institut f\"ur Festk\"orpertheorie, Universit\"at
  M\"unster, 48149 M\"unster, Germany}
\author{Doris~E. Reiter}
\affiliation{Institut f\"ur Festk\"orpertheorie, Universit\"at
  M\"unster, 48149 M\"unster, Germany}
\author{Tilmann Kuhn}
\affiliation{Institut f\"ur Festk\"orpertheorie, Universit\"at
  M\"unster, 48149 M\"unster, Germany}
\author{Martin Gl\"assl}
\affiliation{Institut f\"ur Theoretische Physik III, Universit\"at
  Bayreuth, 95440 Bayreuth, Germany}
\author{Vollrath Martin Axt}
\affiliation{Institut f\"ur Theoretische Physik III, Universit\"at
  Bayreuth, 95440 Bayreuth, Germany}
\author{Anna Grodecka--Grad}
\affiliation{QUANTOP, Danish National Research Foundation Center for
  Quantum Optics, Niels Bohr Institute, University of Copenhagen,
  DK-2100 Copenhagen \O, Denmark}  
\author{Pawe{\l} Machnikowski}
 \email{Pawel.Machnikowski@pwr.wroc.pl} 
\affiliation{Institute of Physics, Wroc{\l}aw University of
Technology, 50-370 Wroc{\l}aw, Poland}

\begin{abstract}

We study the evolution of an exciton confined in a quantum dot
adiabatically controlled by a 
frequency-swept (chirped) laser pulse in the presence of
carrier-phonon coupling. We focus on the dynamics induced by a linearly
polarized beam and analyze the decoherence due to phonon-assisted
biexciton generation. We show that if the biexciton state is shifted
down by a few meV, as is typically the case, the resulting decoherence
is strong even at low temperatures. 
As a result, efficient state preparation is restricted to a small
parameter area corresponding to low temperatures, positive chirps and
moderate pulse areas. 
\end{abstract}


\maketitle

\section{Introduction}

The recently demonstrated \cite{simon11,wu11} 
high-fidelity preparation of a single exciton state in a
self-assembled quantum dot (QD) by means of adiabatic evolution induced
by a chirped laser pulse (referred to as \textit{adiabatic rapid
 passage}, ARP) opens new possibilities of charge control in QDs. In
contrast to the more traditional Rabi flops \cite{zrenner02} as well
as to methods based on voltage control 
\cite{stufler06b,michaelis10}, ARP is much less sensitive to the
details of the driving field. In particular, in the ideal case, the
ARP technique ensures 
the full inversion of occupation between the empty dot and exciton states as soon
 as the pulse intensity reaches the threshold for the adiabatic passage. 

However, as QDs are embedded in a semiconductor
crystal matrix, carrier-phonon interactions are usually found to
considerably limit the fidelity of various optical control
schemes. The phonon-induced dephasing process is inevitable in such
systems and leads to loss of information.  
These mechanisms have been investigated theoretically
\cite{forstner03,krugel05,glaessl11} and have been confirmed experimentally 
\cite{ramsay10,ramsay10b}. As we have recently shown \cite{luker12},
in the case of the ARP the role of phonon coupling in the state
preparation is essential. 
As the adiabatic evolution follows one of the spectral branches
corresponding to dressed exciton states, carrier phonon coupling leads
to transitions to the other branch. Depending on whether the upper or
lower branch is followed (which, in turn, depends on the direction of
the chirp), the transition is assisted by phonon emission or
absorption, respectively. As a consequence, at low temperatures the 
phonon-induced decoherence leads to a strong asymmetry 
in the final occupation of the exciton state depending on the sign of the 
chirp \cite{luker12}. While the analysis performed in Ref.~\onlinecite{luker12} was
based on a two-level model, corresponding to the excitation with a
circularly polarized 
laser pulse, in actual experiments linearly polarized beams have been
used \cite{simon11}. In this case, the selection rules allow the
coupling 
to the biexciton state and phonon-assisted biexciton generation
\cite{findeis00} may occur even at low temperatures, as the biexciton state is typically
shifted to lower energies. 

In this paper, we study the evolution of the biexciton system driven
by a linearly polarized chirped pulse in the regime of the ARP in the presence of
the coupling to acoustic phonons.  We show that including the
biexciton state, which becomes relevant for linearly 
polarized laser pulses \cite{simon11}, indeed opens a new decoherence
path related to phonon-assisted transitions to the biexciton state. 
As a result, the fidelity of exciton state preparation drops down
considerably in particular in most of the parameter areas where almost
ideal evolution was predicted by the two-level model. Thus,
high-fidelity control becomes restricted to the range of moderate
pulse areas for positive chirps and  at low temperatures only, unless
circular polarized pulses are used to suppress biexcitonic
excitations. 

The paper is organized as follows. In Sec.~\ref{sec:model}, we
describe the system and define the model used in our
study. Sec.~\ref{sec:simul} introduces our method of simulation of the
open system dynamics. The results are presented in
Sec.~\ref{sec:results}. Finally, Sec.~\ref{sec:concl} concludes the paper.

\section{Model}
\label{sec:model}

The biexciton system in the QD, restricted to the lowest bright states,
is modeled as a four-level system with $|0\rangle$ representing the
empty dot, $| X \rangle$,$| Y \rangle$ being the two exciton states with different
linear polarizations and $|B \rangle$ denoting the biexciton
state. The corresponding Hamiltonian is
\begin{displaymath}
H_{X}=E_{0}\left( |X \rl X | + | Y \rl Y | \right)
+(2E_{0}+E_{\mathrm{B}})|B\rl B|,
\end{displaymath}
where $E_{0}$ is the exciton transition energy, $E_{\mathrm{B}}$ is
the biexciton shift, and we neglect the fine structure splitting which
is irrelevant on the picosecond time scales considered here.

This charge system is driven by a chirped Gaussian pulse with the
original envelope
\begin{displaymath}
\Omega_{0}(t)=\frac{\Theta}{\sqrt{2\pi}\tau_{0}}
\exp \left (-\frac{t^{2}}{2\tau_{0}^{2}} \right),
\end{displaymath}
where $\Theta$ is the original pulse area (before chirping) and
$\tau_{0}$ is the initial duration 
(pulse length). The central frequency $\omega_{0}$
is chosen at resonance with the single exciton transition.
We assume that the linear chirp is effected
by passing this pulse 
through a Gaussian chirp filter \cite{saleh07} with the chirp
coefficient $\alpha$. This results in the
chirped pulse envelope
\begin{displaymath}
\Omega(t)=\frac{\Theta}{\sqrt{2\pi\tau_{0}\tau}}
\exp\left(-\frac{t^{2}}{2\tau^{2}}\right),
\end{displaymath}
with the chirped pulse length
$\tau=(\alpha^{2}/\tau_{0}^{2}+\tau_{0}^{2})^{1/2}$,
and the frequency chirp rate 
$a=\alpha/(\alpha^{2}+\tau_{0}^{4})$.
The Hamiltonian describing the coupling to the chirped pulse in the
rotating wave approximation is
\begin{displaymath}
H_{\mathrm{las}}=\frac{\hbar \Omega(t)}{2} 
\left( |0 \rl X | + |X \rl B|  \right)
 e^{i\omega_{0}t+iat^{2}/2} +\mathrm{h.c.}
\end{displaymath}
Thus, for one linear polarization, the coupling is 
$|0\rangle\leftrightarrow|X\rangle\leftrightarrow|B\rangle$ and the
other single-exciton state $|Y \rangle$ is decoupled (it is not
coupled by phonon transitions, either\cite{glaessl12}). Therefore, only
a three-level model is needed\cite{schmidgall10}.  

\begin{figure}[tb]
\begin{center}
\includegraphics[width=89mm]{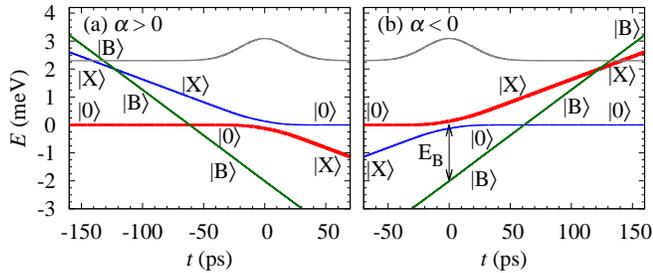}
\end{center}
\caption{\label{fig:branches}(Color online) The adiabatic spectral
  branches of the optically driven QD system for
  $\Theta=2\pi$, $\tau_{0}=2$~ps, $|\alpha|=40$~ps$^{2}$ and 
  $E_{\mathrm{B}}=-2$~meV: (a) positive chirp; (b) negative chirp.
 The evolution starting from the ground
  state follows the red line. The gray solid line represents the pulse
  envelope $\Omega(t)$ which has been shifted upwards for clarity.} 
\end{figure}

The idea of the adiabatic passage \cite{allen75} is to drive the
system state adiabatically along
one of the spectral branches corresponding to the instantaneous
eigenstates of the Hamiltonian
$H_{0}=H_{\mathrm{X}}+H_{\mathrm{las}}$.
These instantaneous eigenstates are shown in
Fig.~\ref{fig:branches} for the biexciton shift
$E_{\mathrm{B}}=-2.0$~meV.
The branch followed by the system state is plotted with a thicker
line. In Fig.~\ref{fig:branches}(a), we assume a positive chirp
($\alpha>0$), while in Fig.~\ref{fig:branches}(b) a negative chirp is
assumed. For pulse intensities above the
ARP threshold, the central anticrossing at $t=0$ (between the vacuum
and exciton states) is passed adiabatically along a single spectral
branch. On the contrary, the two other anticrossings away from $t=0$
(exciton-biexciton and vacuum biexciton) are met when the
pulse is already very weak, hence they are very narrow (in the
case shown here, the widths of both of these anticrossings are below
10~$\mu$eV, unresolved in Fig.~\ref{fig:branches}). At such narrow
anticrossings, the sweep rate is very high
compared to the distance between the two levels, which leads to a
nearly purely non-adiabatic evolution: these anticrossings are almost
completely crossed by the evolving system state, that is, the system
nearly completely jumps to the other branch (as shown by color coding
of the lines in Fig.~\ref{fig:branches}). As a result, the charge
configuration does not change here, as opposed to the central
anticrossing.

For the parameter range corresponding to the experiments
\cite{simon11,wu11} excitation of  
optical phonon modes is excluded by the spectral properties of the system. 
In consequence, the coupling to acoustic phonons can be expected to
dominate as a source of decoherence.    
The carrier-phonon interaction is described in terms of the usual
independent boson Hamiltonian,
$H_{\mathrm{int}} =S\otimes R$, 
with
\begin{displaymath}
S = |X \rl X|+|Y \rl Y| +2 |B \rl B|
\end{displaymath}
and
\begin{displaymath}
R= \sum_{\kk} g_{\kk} \bk + \mathrm{h.c.},
\end{displaymath}
where $\bk$ are phonon annihilation operators. We assume that the
biexcitonic and excitonic wave functions can be factorized 
into products of single particle wave functions which are the
same for both exciton and biexciton. This is well justified in 
the strong confinement limit where the wave functions are mostly determined
by the confinement potentials. The coupling constants, accounting
for the deformation potential interaction, can then be written as
\begin{displaymath}
g_{\kk} = \sqrt{ \frac{\hbar k}{2 \varrho v c} }
[D_{\mathrm{e}}\mathcal{F}_{\mathrm{e}}(\kk)-D_{\mathrm{h}}\mathcal{F}_{\mathrm{h}}(\kk)],
\end{displaymath}
where
$v$ is the normalization volume, $\varrho$ is the crystal density, $c$ is the
speed of sound, $D_{e(h)}$ is the deformation potential constant for
an electron (hole) and 
\begin{displaymath}
\mathcal{F}_{e(h)}(\kk) = \int d^{3}r |\psi_{e(h)}(\bm{r})|^{2} e^{i\bm{k}\cdot\bm{r}}
\end{displaymath}
are the form factors for electron (hole) wave function
$\psi_{e(h)}(\bm{r})$. Finally, the phonon reservoir is described by the
Hamiltonian 
\begin{displaymath}
H_{\mathrm{ph}}=\sumk\hbar\wk\bkd\bk,
\end{displaymath}
where $\wk=ck$ are
the phonon frequencies.

The system is conveniently described in the non-uniformly rotating
frame defined by the unitary transformation $e^{iA(t)}$, where
\begin{displaymath}
A(t)=\left(\omega_{0}t+\frac{1}{2}at^{2}\right)
(|X\rl X| + |Y \rl Y| +2|B \rl B |).
\end{displaymath}
The
resulting Hamiltonian is 
$H=H_{X}'+H_{\mathrm{las}}'+H_{\mathrm{int}}+H_{\mathrm{ph}}$,
with
\begin{align*}
H_{X}'&
=e^{iA(t)}H_{X} e^{-iA(t)}
-\hbar \dot{A}(t) \\
&= \hbar\Delta(t)\left( |X \rl X| + |Y \rl Y|\right) 
+\left(2\hbar\Delta(t)+E_{\mathrm{B}}\right)|B \rl B | ,
\end{align*}
where $\Delta(t)=-at$, and
\begin{align*}
H_{\mathrm{las}}'&
=e^{iA(t)}H_{\mathrm{las}} e^{-iA(t)} \\
&=\frac{\hbar \Omega(t)}{2} 
\left( |0 \rl X | + |X \rl B|  \right)
+\mathrm{h.c.}
\end{align*}

\section{Method of simulation}
\label{sec:simul}

To simplify the numerical simulation, we perform the unitary transformation
defined by the operator
\begin{displaymath}
\mathbb{W}=|0\rl 0|+\left(|X \rl X| + |Y \rl Y| \right) W
+|B \rl B | W^{2},
\end{displaymath}
where 
\begin{displaymath}
W=\exp\left[ \sumk \frac{\gk}{\wk}( \bk-\bmkd ) \right].
\end{displaymath}
Upon this transformation and
expanding to the leading order in the coupling constants, the
Hamiltonian may be written as 
\begin{displaymath}
\tilde{H}=\mathbb{W}H\mathbb{W}^{\dag}=H_{1}+H_{\mathrm{ph}}+V,
\end{displaymath}
where $H_{1} = H_{X}'+wH_{\mathrm{las}}'$
and
\begin{displaymath}
V=\frac{\hbar \Omega(t)}{2} 
\left( |0 \rl X | + |X \rl B|  \right)W^{\dag}
+\mathrm{h.c.}
\end{displaymath}
Here
\begin{displaymath}
w=1-\frac{1}{2}\sumk\left|\frac{\gk}{\wk}\right|^{2}
\end{displaymath}
accounts for the phonon-induced renormalization of the pulse amplitude
in the slow driving limit \cite{krugel05,ramsay10b}. We assume that
the central frequency of the driving pulse is corrected for the
phonon-induced energy shifts. 
The evolution of the reduced density matrix of the charge subsystem
is then found by numerically solving the lowest-order time-convolutionless (TCL)
evolution equation \cite{breuer02,machnikowski08c}
\begin{displaymath}
\dot{\rho}(t)=
-\int_{0}^{t}d\tau\tr_{\mathrm{ph}}\left[ 
V(t),\left[ V(\tau),\rho(t)\otimes\rho_{\mathrm{ph}} 
\right]  \right],
\end{displaymath}
where $V(t)$ is the interaction Hamiltonian $V$
in the interaction picture with respect to $H_{1}+H_{\mathrm{ph}}$,
$\rho_{\mathrm{ph}}$ is the phonon density matrix at the thermal
equilibrium, and $\tr_{\mathrm{ph}}$ denotes the partial trace with
respect to phonon states.  

\begin{figure}[t]
\begin{center}
\includegraphics[width=85mm]{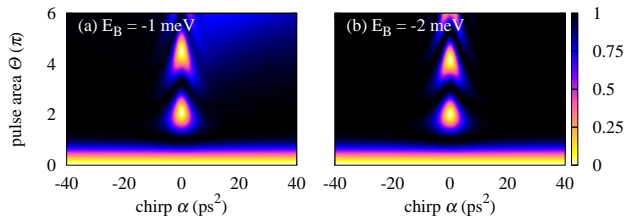}
\end{center}
\caption{\label{fig:maps-bf}(Color online) The final occupation of the
  single-exciton state (color coded) as
  a function of the original pulse area 
$\Theta$ and the chirp $\alpha$ in the absence of coupling to phonons:
(a) for 
$E_{\mathrm{B}}=-1.0$~meV (b) 
for  $E_{\mathrm{B}}=-2.0$~meV.} 
\end{figure}

\begin{figure}[t]
\begin{center}
\includegraphics[width=85mm]{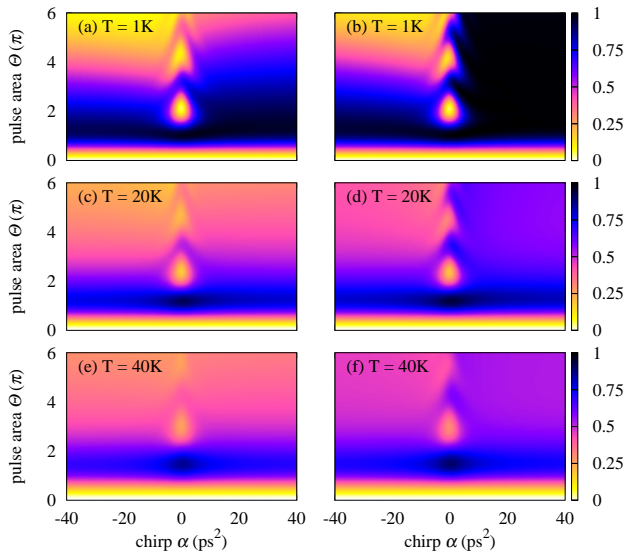}
\end{center}
\caption{\label{fig:maps}(Color online) The final occupation of the
  single-exciton state (color coded) as a function of the original pulse area
$\Theta$ and the chirp $\alpha$  with phonon effects included: (a,c,e)
for  $E_{\mathrm{B}}=-2.0$~meV (b,d,f) for  $E_{\mathrm{B}}=-8.0$~meV.}  
\end{figure}

\begin{figure}[t]
\begin{center}
\includegraphics[width=85mm]{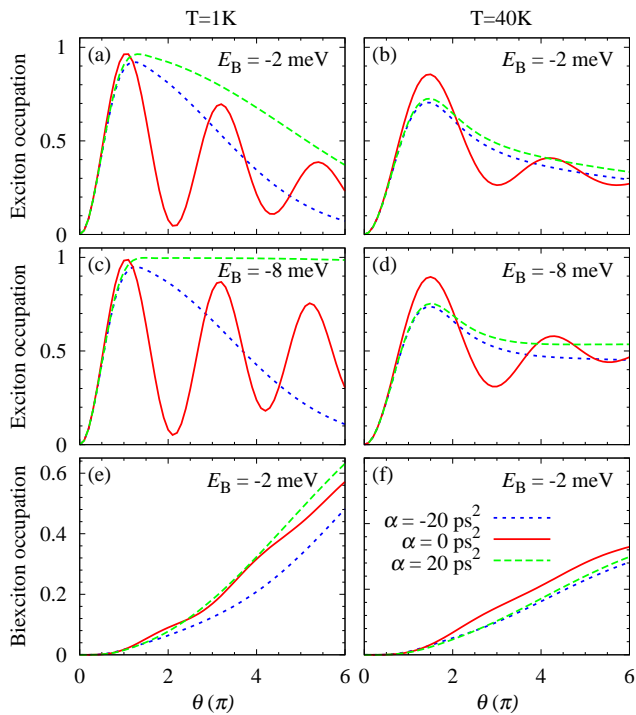}
\end{center}
\caption{\label{fig:cs}(Color online) The final occupation of the
  single-exciton state as a function of the original pulse area 
$\Theta$ at two different temperatures:  (a,b) for
$E_{\mathrm{B}}=-2.0$~meV; (c,d) for 
$E_{\mathrm{B}}=-8.0$~meV. Blue dotted line shows the results with
chirp $\alpha = -20$~ps$^2$, the red solid line represents $\alpha
= 0$, while the green dashed line corresponds to $\alpha =
20$~ps$^2$. The final occupation of the biexciton state for
$E_{\mathrm{B}}=-2.0$~meV is shown in (e,f). }  
\end{figure}

\begin{figure}[t]
\begin{center}
\includegraphics[width=85mm]{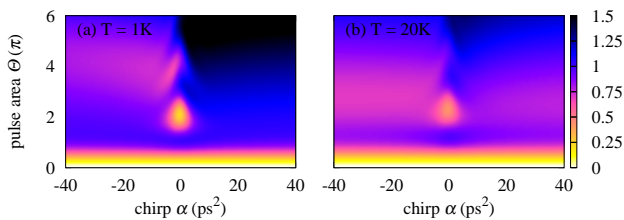}
\end{center}
\caption{\label{fig:maps-total}(Color online) The total number of
  excitons (color coded) as a function of the original pulse area 
$\Theta$ and the chirp $\alpha$ for $E_{\mathrm{B}}=-2.0$~meV (a) at
$T=1$~K and (b) at $T=20$~K.}  
\end{figure}

\section{Results}
\label{sec:results}

With the simulation method outlined in Sec.~\ref{sec:simul}, we study
the exciton dynamics focusing on the possibility of phonon-assisted
transitions to the biexciton state, when the system is excited with a
linearly polarized laser pulse.
In our simulations, we assume $\varrho=5360$~kg/m$^{3}$, $c=5110$~m/s,  
$D_{\mathrm{e}}=7$~eV and $D_{\mathrm{h}}=-3.5$~eV. The electron wave functions are
assumed to be Gaussians with the extension $l_{e}=4.5$~nm in the $xy$
plane and $l_{e,z}=1$~nm along the growth direction. In the case of
holes, we take $l_{h}= 0.87 l_{e}$ and $l_{h,z}= 0.87 l_{e,z}$
respectively.   
The pulse length (before chirping) is taken to be $\tau_{0}=2$~ps.

First, we need to assess the
degree of perturbation to the adiabatic evolution which stems from the
additional level crossings with the biexciton (see
Fig.~\ref{fig:branches}) and is not related to phonons. To this end, 
in Fig.~\ref{fig:maps-bf} we show the single-exciton occupation
without carrier-phonon coupling for $E_{\mathrm{B}}=-1.0$~meV and
$E_{\mathrm{B}}=-2.0$~meV. The results show a reduction of the final
occupation of the single exciton state at higher pulse areas for a
very small biexciton
shift ($E_{\mathrm{B}}=-1.0$~meV)  and positive chirps. We have found
that under these conditions, a biexciton
occupation on the order of $0.1$ occurs (not shown). On the other hand, for
the still small value of 
$E_{\mathrm{B}}=-2.0$~meV, the evolution is nearly unperturbed by the
presence of the biexciton state. 
Thus we conclude that for $|E_{\mathrm{B}}|\gtrsim 2$~meV no direct
population of the biexciton occurs with the pulses studied here, as
the driving of the exciton to biexciton transition 
is too off-resonant.

In the next step, we include phonons and consider two cases:
a) the biexciton state is shifted by a typical energy
($E_{\mathrm{B}}=-2.0$~meV) below the single-exciton state, b) for
comparison, the
biexciton is shifted by a very large (and rarely realized) energy
($E_{\mathrm{B}}=-8.0$~meV). In the first case, one expects that
phonon-induced transitions to the biexciton state are efficient. In
the 
second case, due to the large energy separation between the states (beyond
the cut-off frequency of the carrier-phonon coupling), the biexciton
generation is expected to be negligible and the system can effectively
be treated as a two-level system.  
The simulated evolution of the single exciton occupation for both cases is
shown in Fig.~\ref{fig:maps}. The left panels (Fig.~\ref{fig:maps}(a,c,e)) contain
results for $E_{\mathrm{B}}=-2.0$~meV, and the right panels
(Fig.~\ref{fig:maps}(b,d,f)) for 
$E_{\mathrm{B}}=-8.0$~meV. 
The final occupation is calculated as a function of the pulse area
$\Theta$ and the chirp rate $\alpha$ at three temperatures ($T=1$,
$20$, and $40$~K). 

Let us start the discussion for the case of the large biexciton shift
$E_{\mathrm{B}}=-8.0$~meV and low temperature $T=1$~K shown in
Fig.~\ref{fig:maps}(b). As expected, the results from the two-level
model \cite{luker12} are retrieved and a strong asymmetry in the final
occupation with respect to the chirp is seen. In the two-level model,
for negative chirp the evolution follows the upper branch and phonon
emission leads to damping, while for positive chirps the lower branch
is followed and since phonon absorption is almost absent at 1~K a full 
exciton occupation is reached. In contrast, in Fig.~\ref{fig:maps}(a),
for the small biexciton shift $E_{\mathrm{B}}=-2.0$~meV, the influence 
of the biexciton state can be clearly seen. In particular, for
positive chirps a decoherence at high pulse areas is visible due to
the opening of the new decoherence channel consisting in transitions
to the biexciton state (phonon-assisted biexciton generation).  As can
be seen in 
Fig.~\ref{fig:branches}, for a negative biexciton shift (positive
biexciton binding energy), this process corresponds to phonon emission
no matter which adiabatic branch is followed, hence it is effective
irrespective of the sign of the chirp 
even at low temperatures. Also
for negative chirps the damping has become more effective. Thus, only
for moderate pulse areas around $\Theta=2\pi$ a high-fidelity
preparation of the single exciton state is possible. 

When considering higher temperatures, as seen in Fig.~\ref{fig:maps}(c-f),
phonon absorption becomes more likely and now also for positive chirps
decoherence due to transitions between the main branches is
stronger. Hence, considerable phonon-induced perturbation is seen even
in the case of strongly detuned biexciton
(Fig.~\ref{fig:maps}(d,f)). In consequence, the additional effect of the
biexciton shift is relatively weak.

For a more quantitative analysis we show the final occupation as
a function of pulse area for different chirps $\alpha=-20,0,$ and
$20$~ps$^2$ in Fig.~\ref{fig:cs}(a-d) for $E_{\mathrm{B}}=-2.0$~meV
(Fig.~\ref{fig:cs}(a,b)) and for $E_{\mathrm{B}}=-8.0$~meV
(Fig.~\ref{fig:cs}(c,d)). The most remarkable difference for the two
biexciton shifts is again seen for positive chirp $\alpha=20$~ps$^2$
at $T=1$~K. While in Fig.~\ref{fig:cs}(a), for the small biexciton
shift, a strong damping of the exciton occupation is found, in
Fig.~\ref{fig:cs}(c), for the large biexciton shift, an exciton
occupation of one is reached for all pulse areas above the adiabatic
threshold. Also the decoherence for the negative chirp
$\alpha=-20$~ps$^2$ and damping of the Rabi oscillations at zero chirp
are 
stronger due to the phonon-assisted biexciton generation. For higher
temperature at $T=40$~K in Fig.~\ref{fig:cs}(b,d) the difference
between small and large biexciton shift is less pronounced.  

The strong decoherence for the small biexciton shift
$E_{\mathrm{B}}=-2.0$~meV is accompanied by an occupation of the
biexciton state as shown in Fig.~\ref{fig:cs}(e,f) confirming our
previous statements. 

Since in some experiments \cite{wu11} a photocurrent detection scheme is
used, in which the detection signal is proportional to the total
number of photocreated excitons, we have calculated also the total
number of excitons in the system, $N_{\mathrm{total}} = 
N_{\mathrm{X}}+2 N_{\mathrm{B}}$, where
$N_{\mathrm{X}}$ and $N_{\mathrm{B}}$ 
are the occupation of the single- and biexciton state
respectively.   
The results are shown in Fig.~\ref{fig:maps-total} for $E_{B}=-2$~meV
at $T=1$~K (Fig.~\ref{fig:maps-total}(a)) and $T=20$~K
(Fig.~\ref{fig:maps-total}(b)). One can see that at a certain level of
phonon-assisted biexciton generation the result exceeds 1, which has
no sense if interpreted in terms of the two-level model. Moreover, it
is clear that even below this limit, the biexciton contribution leads
to an overestimated occupation of the single exciton state. 

\section{Conclusions}
\label{sec:concl}

In summary, we have studied the open system evolution of a QD
adiabatically controlled by a chirped linearly polarized laser pulse. 
We have shown that the presence of a biexciton state, which becomes
optically coupled for linearly 
polarized laser pulses \cite{simon11}, opens new decoherence paths
related to phonon-assisted biexciton generation. If the biexciton
state is shifted down by $1$ or $2$~meV it becomes accessible via
phonon emission processes and the resulting decoherence is strong even
at low temperatures. This reduces the achieved occupation of the
single exciton state in the parameters range where the chirped pulse
control was very efficient in a model without the biexciton
state. Since the phonon-assisted biexciton generation increases with
the pulse area moderate pulse areas are optimal for the state
preparation. On the other hand, if the biexciton shift is larger, the
two-level evolution is essentially recovered.  

\acknowledgments

This work was supported in part by a Research Group
Linkage Project of the Alexander von Humboldt Foundation
and by the TEAM programme of the Foundation for Polish
Science, cofinanced from the European Regional Development
Fund.


\begin{thebibliography}{10}

\bibitem{simon11}
C.-M. Simon, T. Belhadj, B. Chatel, T. Amand, P. Renucci, A. Lemaitre, O.
  Krebs, P.~A. Dalgarno, R.~J. Warburton, X. Marie, and B. Urbaszek, Phys. Rev.
  Lett. {\bf 106},  166801  (2011).

\bibitem{wu11}
Y. Wu, I. M. Piper, M. Ediger, P. Brereton, E. R. Schmidgall, P. R. Eastham, M. Hugues,
  M. Hopkinson, and R. T. Phillips, Phys. Rev. Lett. {\bf 106},  067401  (2011).

\bibitem{zrenner02}
A. Zrenner, E. Beham, S. Stufler, F. Findeis, M. Bichler, and G. Abstreiter,
  Nature {\bf 418},  612  (2002).

\bibitem{stufler06b}
S. Stufler, P. Ester, A. Zrenner, and M. Bichler, Phys. Rev. Lett. {\bf 96},
  037402  (2006).

\bibitem{michaelis10}
S. {Michaelis De Vasconcellos}, S. Gordon, M. Bichler, T. Meier, and A.
  Zrenner, Nature Photonics {\bf 4},  545  (2010).

\bibitem{forstner03}
J. F{\"o}rstner, C. Weber, J. Danckwerts, and A. Knorr, Phys. Rev. Lett. {\bf
  91},  127401  (2003).

\bibitem{krugel05}
A. Kr{\"u}gel, V.~M. Axt, T. Kuhn, P. Machnikowski, and A. Vagov, Appl. Phys. B
  {\bf 81},  897  (2005).

\bibitem{glaessl11}
M. Gl\"assl, A. Vagov, S. L\"uker, D.~E. Reiter, M.~D. Croitoru, P.
  Machnikowski, V.~M. Axt, and T. Kuhn, Phys. Rev. B {\bf 84},  195311  (2011).

\bibitem{ramsay10}
A.~J. Ramsay, A. Gopal, E.~M. Gauger, A. Nazir, B.~W. Lovett, A.~M. Fox, and
  M.~S. Skolnick, Phys. Rev. Lett. {\bf 104},  017402  (2010).

\bibitem{ramsay10b}
A.~J. Ramsay, T.~M. Godden, S.~J. Boyle, E.~M. Gauger, A. Nazir, B.~W. Lovett,
  A.~M. Fox, and M.~S. Skolnick, Phys. Rev. Lett. {\bf 105},  177402  (2010).

\bibitem{luker12}
S. L\"uker, K. Gawarecki, D.~E. Reiter, A. Grodecka-Grad, V.~M. Axt, P.
  Machnikowski, and T. Kuhn, Phys. Rev. B {\bf 85},  121302(R)  (2012).

\bibitem{findeis00}
F. Findeis, A. Zrenner, G. B{\"o}hm, and G. Abstreiter, Phys. Rev. B {\bf 61},
  R10579  (2000).

\bibitem{saleh07}
B. Saleh and M. Teich, {\em {Fundamentals of photonics}}, {\em Wiley series in
  pure and applied optics} (Wiley-Interscience, New York, 2007).

\bibitem{glaessl12}
M. Gl\"assl, M.~D. Croitoru, A. Vagov, V.~M. Axt, and T. Kuhn, Phys. Rev. B
  {\bf 85},  195306  (2012).

\bibitem{schmidgall10}
E.~R. Schmidgall, P.~R. Eastham, and R.~T. Phillips, Phys. Rev. B {\bf 81},
  195306  (2010).

\bibitem{allen75}
L. Allen and J.~H. Eberly, {\em Optical resonance and two-level atoms} (Wiley,
  New York, 1975).

\bibitem{breuer02}
H.-P. Breuer and F. Petruccione, {\em The Theory of Open Quantum Systems}
  (Oxford University Press, Oxford, 2002).

\bibitem{machnikowski08c}
P. Machnikowski, Phys. Rev. B {\bf 78},  195320  (2008).

\end{thebibliography}

\end{document}